# Zero-field propagation of spin waves in waveguides prepared by focused ion beam direct writing

Lukáš Flajšman[1,*], Kai Wagner[2], Marek Vaňatka[1], Jonáš Gloss[3], Viola Křižáková[4], Michael Schmid[3], Helmut Schultheiss[2] and Michal Urbánek[1,4*]

[1] *CEITEC BUT, Brno University of Technology, Brno, Czech Republic*
[2] *Institute of Ion Beam Physics and Materials Research, HZDR, Dresden, Germany*
[3] *Institute of Applied Physics, TU Wien, Vienna, Austria*
[4] *Institute of Physical Engineering, Brno University of Technology, Brno, Czech Republic*

[*]E-mail: lukas.flajsman@ceitec.vutbr.cz, michal.urbanek@ceitec.vutbr.cz

Metastable $Fe_{78}Ni_{22}$ thin films are excellent candidates for focused ion beam direct writing of magnonic structures due to their favorable magnetic properties. The focused ion beam transforms the originally nonmagnetic fcc phase into the ferromagnetic bcc phase with an additional control over the direction of uniaxial magnetic in-plane anisotropy and the saturation magnetization. The induced anisotropy allows to stabilize transverse direction of magnetization in narrow waveguides. Therefore, it is possible to propagate spin waves in these waveguides in the favorable Demon-Eshbach geometry without the presence of any external magnetic field.

Nowadays, the vibrant field of magnonics stands on the edge between development of elementary building blocks of magnonic circuitry and envisioned all magnon on-chip devices [1,2]. The magnonic devices, utilizing physics of spin waves, are recognized to have potential in information processing in the frequency range from gigahertz to terahertz. High frequencies, together with low energy of elementary excitations render the magnonic devices suitable for beyond-CMOS computational technologies. Many concepts of future devices used for steering and manipulating spin waves have been presented recently [3-6]. To allow further advances in this field, new types of materials possessing additional means of control over their magnetic properties together with good spin wave propagation are needed.

Here we show, that magnonic waveguides allowing for fast spin-wave propagation at zero magnetic field can be directly written into metastable $Fe_{78}Ni_{22}$ thin films by focused ion beam (FIB). The local dose and scanning strategy controls both the saturation magnetization and the magnetocrystalline anisotropy (direction, type and strength) of irradiated areas. The unique possibilities of this material system allow to overcome the shape anisotropy of long magnonic waveguides and stabilize the magnetization along the waveguide's short side even in the absence of any external magnetic field. This is of great importance for using magnons as information carriers because the geometry where the magnetization is aligned perpendicularly to the propagation direction has the maximum group velocity.

The $Fe_{78}Ni_{22}$ layers grow on Cu(001) substrate in the metastable nonmagnetic fcc phase and can be transformed by ion irradiation into the ferromagnetic bcc phase [7]. Using FIB the fcc->bcc transformation can be precisely controlled which results in unprecedented local control over magnetic properties [8]. This approach removes the need for complicated multi-step lithography processing and allow rapid prototyping of individual structures on the sample with additional possibility to control the magnetic properties of each structure [8-12]. Classical approaches of nanostructuring of magnetic materials such as optical or electron beam lithography combined with lift-off processing [13], wet [14] or dry [15] etching or ion implantation [16,17] do not offer the same control over local material properties as the fabrication processes rely on binary selection of adding or removing the magnetic materials.

The FIB-written waveguides together with a microwave antenna for spin-wave excitation are shown in Fig. 1 a). The waveguides are 30 µm long with nominal widths of 3, 2 and 1.5 µm and are clearly visible in the SEM image.

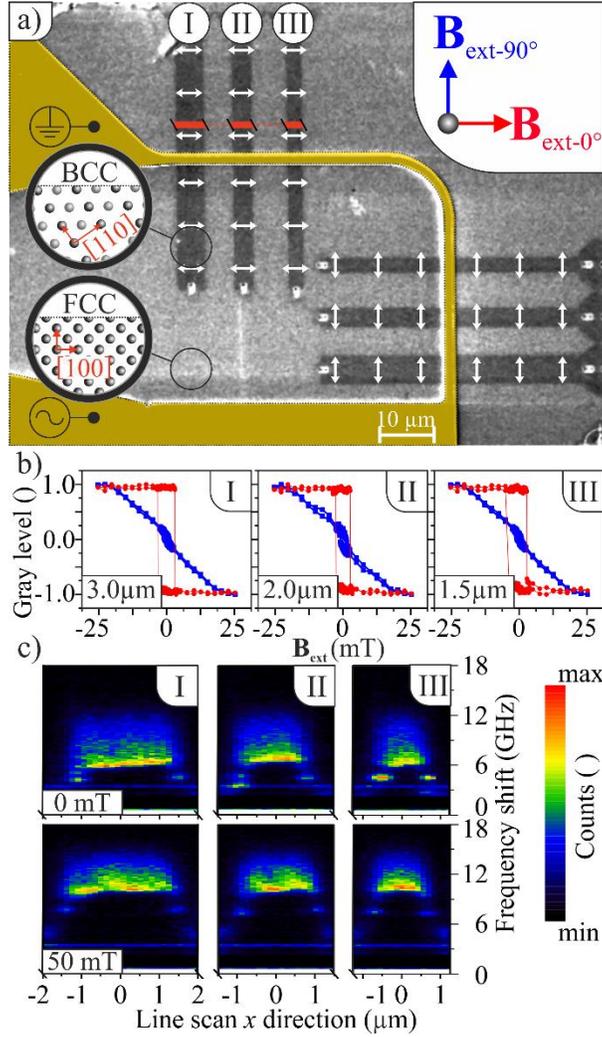

*Fig. 1: a) SEM micrograph of FIB prepared waveguides (bcc waveguides appear dark grey). The crystallographic orientations of the fcc matrix and bcc waveguides are indicated (for the latter, inferred from the magnetic anisotropy, double-headed arrow). The microwave antenna is highlighted by ocher color. b) hysteresis loops of three waveguides for two different orientations of the magnetic field [blue and red arrows showing the respective orientation of the magnetic field are shown in a)]. c) 2D frequency maps of thermal spin-wave spectra measured by micro-focused BLS scanning over the waveguides in the transverse direction along the red lines in (a) in zero external magnetic field (top row) and in the applied magnetic field $B_{ext-0°} = 50$ mT (bottom row). The spatially and magnetic field invariant mode at approx. 3 GHz is a spurious laser mode.*

Prior to the FIB processing the metastable fcc $Fe_{78}Ni_{22}$ thin film of nominal thickness 12 nm was grown under UHV conditions on a Cu(001) single crystal substrate using the procedure described in [7]. After deposition, the sample was transferred to the FIB-SEM microscope for FIB irradiation. We oriented the long axis of the waveguides along the fcc[010] direction of the Cu substrate in order to obtain the highest possible magnetocrystalline anisotropy. To imprint the anisotropy direction perpendicularly to the long axis of the waveguide, we wrote the structures with a single pass of a 30 keV Ga+ ion beam (30 nm spot, beam current of 150 pA and 5 μs dwell time) with the fast scanning direction rotated by 80° from the waveguide's long axis. The resulting ion dose $4 \times 10^{15}$ ions/cm² has given reliable growth conditions for all the waveguides. Nucleation of the bcc structure was facilitated by starting the growth in triangular 5 μm wide region with the doubled ion dose of $8 \times 10^{15}$ ions/cm² (again in single FIB scan).

After the irradiation, the magnetic waveguides were investigated by Kerr magnetometry [18] and Kerr microscopy. When the field $B_{\text{ext}-90°}$ is applied parallel to the long axis of the waveguide, hard-axis hysteresis loops [blue lines in Fig. 1 b)] with effective anisotropy fields [19] in the range of $20 - 24$ mT are observed for all waveguides. When the field $B_{\text{ext}-0°}$ is applied perpendicular to the waveguide, easy-axis loops [red lines in Fig. 1 b)] with coercive fields in the range of $4 - 8$ mT are observed.

Thermal spin-wave spectra obtained by micro-focused Brillouin light scattering microscopy [20] are shown in Fig. 1 c) for zero magnetic field and for the external field $B_{\text{ext}-0°} = 50$ mT. The signal is proportional to the density of spin-wave states at the detection frequency ($y$-axis) and shows pronounced spin-wave spectral intensity localized solely in the areas irradiated by the FIB. The bandgap of the spin-wave band structure for the middle of the waveguide ($x = 0$ μm) can be clearly seen as a sudden increase in the density of spin-wave states for frequencies higher than approx. 6 GHz in zero field and at approx. 10 GHz in the external field of 50 mT.

All three waveguides show qualitatively the same behavior. The data also reveals the local change in spin-wave spectra towards the sides of the waveguide. Here localized low frequency modes appear at approx. 4 GHz in zero field and at approx. 8 GHz in the external field of 50 mT. This is a clear indication of the transverse orientation of the magnetization and its inherent demagnetizing field leading to lower effective fields at the waveguide edges and thus directly resulting in the localized spin-wave edge modes [21,22]. The overall analysis is depicted for three vertically oriented waveguides only as the behavior of the horizontally oriented waveguides show qualitatively same behavior (the anisotropy is again imprinted perpendicular to the long axis of the waveguides). This also demonstrate unique potential of our approach when compared to other less versatile approaches or materials with global magnetocrystalline anisotropy.

In the following experiments, we extract the magnetic field-dependence of the spin-wave dispersion relation. By fitting the measured dispersion, we were able to obtain the full set of magneto-dynamic parameters of the material. The μBLS does not directly sense the phase of the detected spin waves and thus it does not allow to determine the wavelength $\lambda$ (or equivalently the wave propagation vector $\vec{k}$) of the spin waves. In order to extract the wave-vector information we employed the phase-resolved µBLS technique [20,23]. The method directly reveals the spatial profile of the spin-wave phase by letting the scattered photons interfere with a reference spatially invariant signal created by an electro-optic modulator (EOM).

We recorded the interference signal along the 1.5 µm wide waveguide ($x = 0$ µm) with a step size of 120 nm from the edge of the exciting antenna up to 7 µm distance at the microwave frequency of 10.2 GHz. The phase-resolved measurements are shown in Fig. 2.

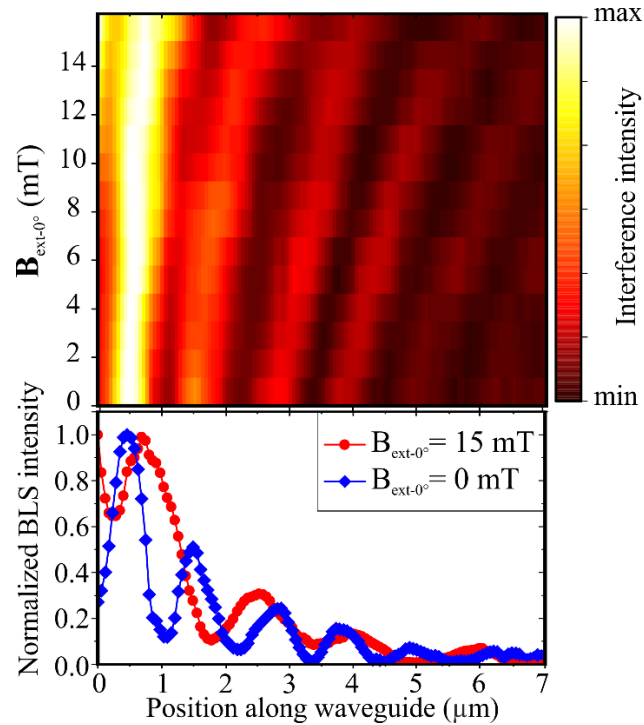

*Fig. 2: Phase-resolved BLS microscopy interference intensity map for various external magnetic fields at fixed excitation frequency of 10.2 GHz. Individual linescans have been normalized and space-invariant background was subtracted. The bottom graph shows line profiles extracted from the intensity map for 15 mT (red line) and for zero external magnetic field (blue line).*

We measured the BLS interference linescans in zero and applied external magnetic field (perpendicular to the waveguide long axis) up to 15 mT (1 mT step). The BLS interference intensity map shows gradual transition of the spin-wave wavelength from the lowest value found at zero field up to the longest wavelengths at 15 mT. This is additional evidence for the presence of the Damon-Eshbach geometry even at zero field since otherwise a decrease in wavelength with increasing field would be expected. To extract the spin-wave wavelength, we fitted the measured data with the simple interference model:

$$I(y) = I_{SW}(y) + I_{EOM} + 2\sqrt{I_{SW}(y)I_{EOM}}\cos(\theta(y)), \quad (1)$$

where we assume $I_{SW} = I_{max}e^{-y/L_{att.}}$. The parameters $I_{EOM}$ and $I_{max}$ are EOM and spin-wave maximum intensities, $L_{att.}$ is a spin-wave attenuation length and the total phase difference is $\theta = 2\pi\frac{y}{\lambda_l} + \theta_0$ with $\lambda_l$ representing the longitudinal spin-wave wavelength, $y$ is the distance from the excitation antenna, and $\theta_0$ is an arbitrary phase offset in-between the EOM and the spin-wave excitation. We fitted the experimental data using Eq. (1) with all five parameters unconstrained. The two important fit parameters are the wavelength [shown in Fig. 3 a) as a function of $B_{ext}$] and the attenuation length. The largest attenuation length $L_{att.} = 3.1 \pm 0.4$ µm is measured for zero external magnetic field and the frequency of 10.2 GHz. The spin-wave dispersion at zero field obtained from these fits is shown in Fig. 3 b).

From the dispersion, we can determine the magnetic parameters using the model of Kalinikos and Slavin [24], while taking into account the finite width of our waveguides by assuming the effective boundary conditions as described by Guslienko et al. [25]. The effective dipolar conditions yield an effective width $w_{eff} = wd/(d-2)$ of the waveguide, determined by the geometric width $w$ and the pinning parameter $d = 2\pi/(p + 2p\ln(1/p))$. The parameter $p$ is given by $p = t/w$, where $t$ is the thickness of the magnetic material. For the spin wave dispersion model, we assumed that the external magnetic field $B_{ext}$ points in the direction of the magnetic anisotropy, i.e. perpendicularly to the long edge of the waveguide (magnetic anisotropy is introduced in the form of effective magnetic field $B_{ani}$). With all the terms in place the model is following:

$$f^2/\gamma^2 = (|B_{ext}| + |B_{ani}| + \mu_0 M_s P \sin^2(\sphericalangle k) + A_{ex}k^2)(|B_{ext}| + |B_{ani}| - \mu_0 M_s(1-P) + A_{ex}|\vec{k}|^2), \quad (2)$$

where $f$ is the spin-wave frequency and $\gamma$ is the gyromagnetic ratio. $M_s$ is the saturation magnetization and $A_{ex}$ is an exchange stiffness. Since the thickness of the waveguide is small with respect to the other dimensions and the spin-wave wavelength, we only calculate the first branch of spin waves along the thickness, yielding $P = 1 - (1 - \exp(-|\vec{k}|t)/|\vec{k}|t)$. The total propagation vector $|\vec{k}| = \sqrt{k_\parallel^2 + k_{\perp n}^2}$ comprises the longitudinal component $k_\parallel$ (parallel to the long axis of the waveguide) and transverse quantized component $k_{\perp n}$. The angle of the spin wave propagation vector with the long waveguide axis is then $\sphericalangle k = \mathrm{atan}(k_\parallel/k_{\perp n})$. The amplitude of the transverse component $k_{\perp n}$ is calculated from the width quantization condition $k_{\perp n} = n\pi/w_{eff}$ for $n = 1, 2, ....$ In the micro-focused BLS experiment, as we record only the interference pattern along the long waveguide axis, we see only the longitudinal component of the propagation vector $k_\parallel = 2\pi/\lambda_l$.

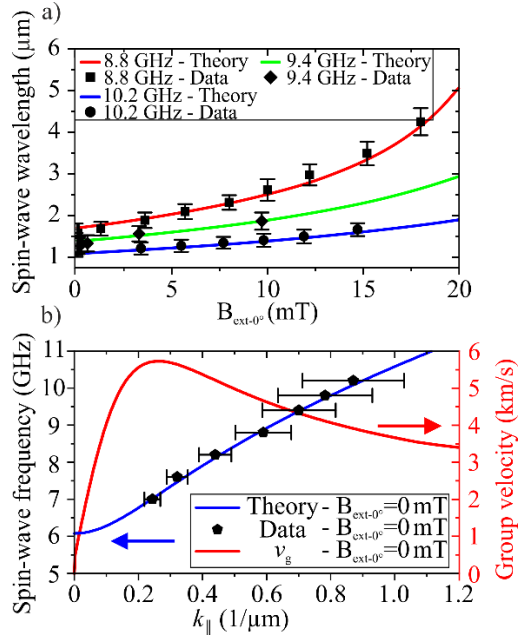

*Fig. 3: a) the dependence of the spin-wave wavelength $\lambda_l$ [extracted from equation (1)] on the external magnetic field for three different RF frequencies fitted with a model given by equation (2). b) experimental and calculated (blue line) spin-wave dispersions together with calculated group velocity (red line) at zero external magnetic field. The error bars have been calculated from the 95% fit confidence bounds.*

We performed the fit using Eq. (2) [considering 95% confidence intervals obtained by fitting of the Eq. (1)] for various width modes (and their linear combinations [26,27]) with one set of unconstrained universal parameters and we observed the total minimal residuals of the fit. The best fit was found solely for single mode of $n = 1$ (in contrast to experiments in e.g. permalloy waveguides [26,27]), with magnetic parameters of $M_s = 1.41 \pm 0.03$ MA/m, $\gamma = 29.3 \pm 0.1$ GHz/T, $B_{ani} = 24 \pm 1$ mT, $t = 9.5 \pm 1.0$ nm, $A_{ex} = 11 \pm 5$ pJ/m and $w = 1.62 \pm 0.05$ µm. The obtained fit parameters lie close to the bulk values of single crystal iron films [28]. The saturation magnetization $M_s = 1.41$ MA/m is expectedly lower then bulk value of iron ($M_s^{Fe} = 1.7$ MA/m). If we consider 22% of nickel ($M_s^{Ni} = 0.51$ MA/m) in our films, we estimate the expected saturation magnetization to $M_s^{FeNi} = 1.45$ MA/m which is very close to the obtained value $M_s = 1.41$ MA/m. The large saturation magnetization together with the Damon-Eshbach geometry result in a high group velocity $v_g = \partial f / \partial k_\parallel$ of the spin waves reaching almost 6 km/s [see Fig. 3 b)]. The anisotropy field resulting from the fit perfectly reproduces the value measured by Kerr microscopy, which further supports the validity of the model. We also see that the film is transformed to the magnetic bcc phase throughout the whole thickness. The fitted thickness is very close to the nominal thickness of 12 nm (the few topmost layers are expected to oxidize when performing ex-situ experiments). This is also supported by the Monte Carlo ion stopping range simulations using the SRIM/TRIM package [29], where more than 90% of the 30 keV Ga+ ions penetrate to the copper substrate. The width of the waveguide $w = 1.62$ μm is slightly larger than the nominal value (1.5 μm). First and presumably the major contribution increasing the width of the waveguide is the finite size and shape of the focused ion beam spot. Furthermore, in our previous work [30] it was shown that the bcc crystallites protrude to the fcc phase slightly further from the ion impact spot and thus they again effectively increase the width of the waveguides (the protrusion length is approx. 50 nm). Both effects effectively create a gradient in the magnetization affecting the dynamic boundary conditions of our waveguides, which differs from the discontinuous boundary conditions found e.g. in structures prepared by classical lithography techniques [26,27]. We performed micromagnetic simulations in mumax$^3$ [31] to study effects of the magnetization gradient at the waveguide edges on the spin-wave dispersion (for the approach see supplementary material [32]). In the simulations we continuously decreased the magnetization from the bulk value to zero in a defined region at the edge of the waveguide. The introduced profile of the magnetization was chosen as an error function as it resembles the convolution of the nominal shape of the waveguide with a FIB spot. As the gradient of the saturation magnetization is introduced it is expected that the effective magnetic field, the major driving force affecting the local spin-wave dispersion, will differ from discontinuous case where the saturation magnetization changes abruptly. The transverse profiles of the saturation magnetization together with simulated effective (internal) magnetic field at zero external magnetic field are shown in Fig. 4 a).

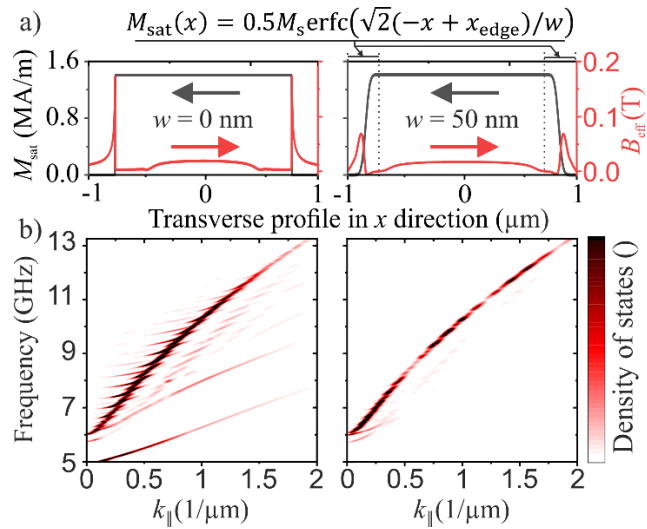

*Fig. 4: a) shows the transverse profiles of the saturation magnetization (grey) and effective magnetic field (red) for discontinuous (left panel) and continuous (right panel) transition of saturation magnetization on the edges of the waveguide at zero external magnetic field for material parameters obtained by the fitting procedure. The continuous transition was implemented as a quasi-continuous modulation of $M_s$ expressed by an equation shown above the plots. Transition width (w) of the complementary error function is designated in each plot. b) dispersion relations of the spin-wave modes for both magnetization profiles extracted from micromagnetic simulations. The density of states (represented by color scale) has been normalized to the maximum value.*

The micromagnetic simulations reveal, that in the case of the infinite gradient in the magnetization leads to abrupt increase of the effective magnetic field ($B_{\text{eff}}$) on the edges of the waveguide. In the case of finite magnetization gradient, the $B_{\text{eff}}$ profile is significantly changed. The maximal value of the $B_{\text{eff}}$ is lower and the central region is significantly broadened, as the lower gradient of magnetization leads to lower and less localized demagnetizing field. The longitudinal spin-wave dispersion extracted from the micromagnetic simulations is plotted in Fig. 4 b). There is a very apparent change in the modal structure of the dispersion when we compare the case with $w = 0$ nm and the case where $w = 50$ nm. As the gradient is introduced to the edge region, the boundary conditions for the dynamic magnetization are altered. The quantization of the modes with higher transverse mode numbers is deteriorated. This leads to less effective excitation of the higher order modes when compared to e.g. fundamental mode (as can be seen from a lower modal density of states). The first waveguide mode does not significantly change even for $w = 200$ nm. This is likely the main mechanism explaining the absence of any higher order waveguide modes seen in the linescans. This statement is further supported by careful analysis of the linescans (see Fig. 2). In additional analysis we readjusted the equation (1) to allow for more spatial frequencies to be detected since we expect from the modal profiles of the analytical model presented by equation (2) to excite multiple odd modes [27,33] at certain frequency by the excitation antenna. The analysis confirmed best agreement in the absence of any higher spatial frequencies.

In conclusion we studied the spin-wave propagation in the waveguides prepared by FIB direct writing into metastable fcc $Fe_{78}Ni_{22}$ thin films. We have shown that in these high-aspect ratio waveguides we can propagate spin waves without the presence of an external magnetic field, and with high group velocities reaching almost 6 km/s. This unique feature comes from the possibility of the local uniaxial magnetic anisotropy control. The spin-wave dispersion relation has been determined by using phase-resolved BLS microscopy and the magnetic properties of the waveguides were extracted. The relatively large saturation magnetization together with high (controllable) magnetic anisotropy render the material suitable for high frequency spin-wave circuits operational even at zero external magnetic field. Moreover, the extracted material properties of the system will allow us to design more complex spin-wave devices by utilizing the possibility to spatially control both the saturation magnetization and the direction of the uniaxial magnetic anisotropy in a single magnetic structure. Our unique approach paves the way towards many other possibilities to develop and study spin-wave propagation in magnetization landscapes that are unattainable in any conventional magnetic system.

Acknowledgements:

This research has been financially supported by the joint project of Grant Agency of the Czech Republic (Project No. 15-34632L) and Austrian Science Fund (Project I 1937-N20) and by the CEITEC Nano+ project (ID CZ.02.1.01/0.0/0.0/16013/0001728). Part of the work was carried out in CEITEC Nano Research Infrastructure (ID LM2015041, MEYS CR, 2016–2019). L.F. was supported by Brno PhD talent scholarship.